\documentclass[nofootinbib, reprint, floatfix]{revtex4-1}

\usepackage{units}
\usepackage{graphicx} 
\usepackage{upgreek}
\usepackage{mathtools}
\usepackage{dcolumn}
\usepackage{bm}
\usepackage[caption=false]{subfig}
\usepackage{color,soul}
\usepackage{tabularx}
\usepackage{xcolor}
\usepackage{framed}

\begin{document}


\title{Induced THz transitions in Rydberg caesium atoms for application in antihydrogen experiments}

\author{M. Vieille-Grosjean}
\author{Z. Mazzotta}\thanks{\emph{Present address:} Advanced Research Center for Nano-lithography (ARCNL), Science Park 106, 1098 XG, The Netherlands}
\author{D. Comparat}
\address{Universit\'e Paris-Saclay, CNRS, Laboratoire Aim\'e Cotton, 91405, Orsay, France}

\author{E. Dimova}\thanks{deceased September 08, 2020}
\address{Bulgarian Academy of Sciences, 72 Tzarigradsko Chauss\'ee Blvd., 1784 Sofia, Bulgaria}

\author{T. Wolz and C. Malbrunot}
\address{Physics Department, CERN, Gen\`eve 23, 1211, Switzerland}

\date{\today}

\begin{abstract}
Antihydrogen atoms are produced at CERN in highly excited Rydberg states. However, precision measurements require anti-atoms in ground state. Whereas  experiments currently rely on spontaneous emission only, simulations have shown that THz light can be used to stimulate the decay towards ground state and thus increase the number of anti-atoms available for measurements. We review different possibilities at hand to generate light in the THz range required for the purpose of stimulated deexcitation. We demonstrate the effect of a blackbody type light source, which however presents drawbacks for this application including strong photoionization.
Further, we report on the first THz transitions in a beam of Rydberg caesium atoms induced by photomixers and conclude with the implications of the results for the antihydrogen case.
\end{abstract}

\maketitle

\section{Introduction}
After years of technical developments, antihydrogen ($\bar{\rm{H}}$) atoms can be regularly produced at CERN's Antiproton Decelerator complex \cite{AlP_Hbar_Accumulation,KUR14,PhysRevLett.108.113002}. This anti-atom is used for stringent tests of the Charge-Parity-Time (CPT) symmetry as well as for the direct measurements of the effect of the Earth's gravitational acceleration on antimatter. For precision measurements towards both of these goals ground-state antihydrogen atoms are needed.

The atoms are mostly synthesized using either a charge exchange (CE) reaction where an excited positronium (Ps) atom (bound state of an electron and a positron) releases its positron to an antiproton or a so-called three-body recombination reaction (3BR) where a large number of positrons and antiprotons are brought together to form antihydrogen.

Both formation mechanisms produce antihydrogen \sloppy atoms in highly excited Rydberg states with so-far best achieved temperatures of $\sim \unit[40]{K}$ \cite{AlP_Hbar_Accumulation} (corresponding to a mean velocity of $\sim\unit[1000]{m/s}$) and in the presence of relatively strong magnetic fields ($\mathcal{O}(\unit[1]{T})$) to confine the charged particles and, in some cases, trap the antihydrogen atoms. Although experimentally not well studied, the antihydrogen atoms formed via 3BR are expected to cover a broad range of principle quantum numbers up to $n \sim 100$ \cite{Gabrielse2002,rsa_Mal_18,ROB08,radics2014scaling,Jonsell_2019}. Highly excited states will be ionized by the electric field present at the edges of the charged clouds so that in general only antihydrogen with $n<50$ can escape the formation region. Via the CE reaction, specific $n \sim 30$ values can be targeted resulting in a narrower spread in $n$ that is mainly determined by the velocity and velocity distribution of the impinging Ps \cite{Krasnicky_2019,2016PhRvA..94b2714K,2012CQGra..29r4009D}. In either case, all ($k,m$) substates are populated where $m$ is the magnetic quantum number and $k$ a labeling index according to the strength of the substate's diamagnetic interaction that becomes in a field-free environment the angular momentum quantum number $l$. The field-free lifetime $\tau_{n,l}$ of the Rydberg states produced 
\begin{equation}
		\tau_{n,l} \approx \left( \frac{n}{30} \right)^3 \left( \frac{l+1/2}{30} \right)^2 \times \unit[2.4]{ms}
\label{eq: RydbergLifetime}
\end{equation}
is of the order of several milliseconds \cite{PhysRevA.31.495} and can be considered a good approximation in the presence of a magnetic field $B \sim \unit[1]{T}$ \cite{Topccu2006}. Given the currently achieved formation temperatures, this results, for experiments that rely on an antihydrogen beam, in a large fraction of atoms remaining in excited states before escaping the formation region which complicates beam formation and hinders in-situ measurements.

In a previous publication \cite{wolz2019stimulated} the stimulation of atomic transitions in (anti-)hydrogen using appropriate light in order to couple the initial population to fast spontaneously decaying levels was studied. Indeed, such techniques allow to increase the ground state fraction within a few microseconds which corresponds to an average flight path of the atoms on the order of centimeters. Different deexcitation schemes, making use of THz light, microwaves and visible lasers were investigated.
Microwave sources and lasers at the wavelengths and intensities identified in \cite{wolz2019stimulated} are commercially available and measurements of single Rydberg-Rydberg transitions have been reported \cite{gallagher_1994}. The simultaneous generation of multiple powerful light frequencies in the high GHz to THz regime however still remains a technical challenge.
After providing some insights into the antihydrogen deexcitation schemes dealt with and clarifying which light intensities and wavelengths are required in section \ref{s: requirements}, we analyse in section \ref{s:source} the suitability of different THz light sources for this purpose. We report in section \ref{s:experiment} on the effect of a broadband lamp and on the first observation of highly selective THz light stimulated population transfer between Rydberg states with a photomixer in a proof-of-principle experiment with a beam of excited caesium atoms.

\section{THz-induced antihydrogen deexcitation and state mixing}
\label{s: requirements}
For both production schemes, CE and 3BR, the idea of stimulated deexcitation of antihydrogen comes down to mixing many initially populated long-lived states and simultaneously driving transitions to fewer short lived-levels from where the spontaneous cascade decay towards the ground state is fast. Relying on a pulsed CE production scheme the initially populated states can be mixed by applying an additional electric field to the already present magnetic one \cite{COM181}. A deexcitation/mixing scheme based on the stimulation of atomic transitions via light in the THz frequency range is thoroughly discussed in \cite{wolz2019stimulated} for the 3BR case. This latter scheme is equally applicable to a pulsed CE production.

When coupling a distribution of $N$ long-lived levels with an average lifetime of $\tau_{\rm{N}}$ to $N'$ levels with an average deexcitation time to ground state of $\tau_{\rm{N'}}^{\rm{GS}} \ll \tau_{\rm{N}}$, the minimum achievable time $t_{\rm{deex}}$ for the entire system to decay to ground state can be approximated by
\begin{equation}
		t_{\rm{deex}} \approx \frac{N}{N'} \times \tau_{N'}^{\rm{GS}}.
\label{eq: DeexTime}
\end{equation}
In (anti)hydrogen, the average decay time of a $n'$-manifold with $N'$ fully mixed states to ground state can be approximated, for low $n'$, by the average lifetime of the manifold:  $\tau_{N'}^{\rm{GS}} \sim \unit[2]{\mu s} \times \left( n'/10 \right)^{4.5}$ \cite{COM181}. Consequently, when coupling some thousands of initially populated Rydberg antihydrogen levels ($n\sim 30$) to a low lying manifold this intrinsic limit would lead to a best deexcitation time towards ground state of roughly a few tens of $\unit{\mu s}$. Within such a time interval the atoms move only by a few tens of $\unit{mm}$ and thus stay close to the formation region from where, once deexcited, a beam can be efficiently formed.

\begin{figure}
\includegraphics[width=1\linewidth]{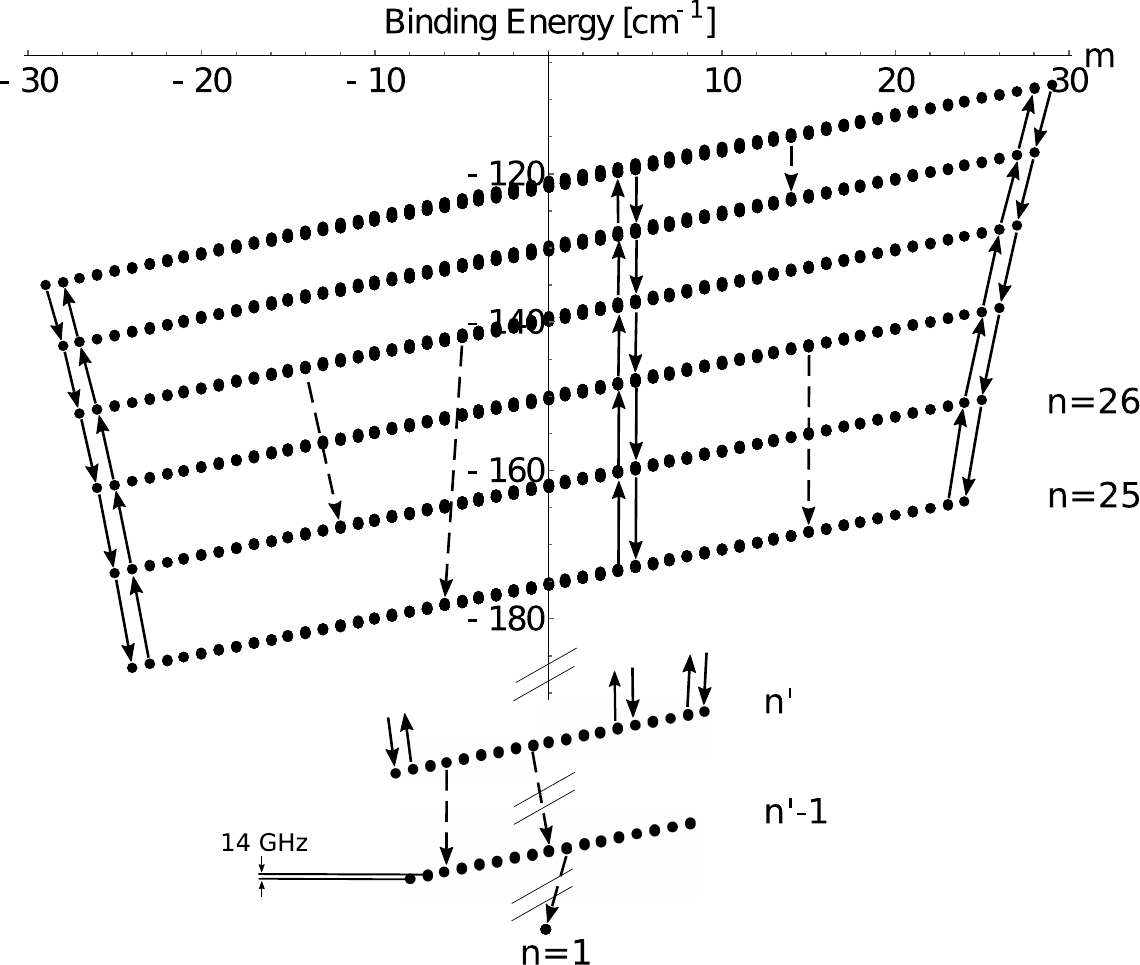}
\caption{Binding energy of (anti-)hydrogen Rydberg states in a $\unit[1]{T}$ magnetic field as a function of the magnetic quantum number $m$. Inter-$n$ manifold transitions in the THz region are indicated by continuous errors. Dashed arrows illustrate some examples of spontaneous transitions. The figure is adapted from Ref.~\cite{wolz2019stimulated}.}
\label{fig: H_Binding_energy}
\end{figure} 

Figure~\ref{fig: H_Binding_energy} shows the binding energy diagram of antihydrogen states in a $\unit[1]{T}$ magnetic field. Recalling the $|\Delta m|=1$ selection rule, it becomes apparent that, especially to address high angular momentum states that are incidentally the longest lived ones, all $\Delta n= -1$ THz-transitions need to be driven to achieve an efficient mixing.
In Ref.~\cite{wolz2019stimulated} the efficiency of stimulating simultaneously all $\Delta n = -1$ inter-$n$ manifold transitions from Rydberg levels $(n,k,m)$ down towards a manifold $n'$ that is rapidly depopulated to ground state by spontaneous emission (in the following referred to as THz deexcitation) is studied. 

For $n=30$ and $n'=5$ it is found that the total (summed over all driven transitions) light intensity necessary is of $> \unit[10]{mW/cm^2}$ covering a frequency range from $\sim \unit[200]{GHz}$ to well within the few $\unit[]{THz}$ region (the frequencies range from over \unit[40]{THz} for $n=6\rightarrow 5$ to \unit[0.26]{THz} for $n=30 \rightarrow 29$).

As an alternative scheme (in the following referred to as THz mixing), it was 
proposed to restrict the THz light to a certain fraction of the initially populated levels in order to mix all ($k,m$) sublevels within, for example, $25 \leq n \leq 35$. Retaining the levels equipopulated allows for a narrowband deexcitation laser to couple the Rydberg state distribution directly to the $n=3$ manifold which decays on a nanosecond timescale. This results in a reduction of the total THz light intensity required by more than an order of magnitude to $\unit[1]{mW/cm^2}$. 

In summary, THz mixing or deexcitation requires the simultaneous generation of multiple light frequencies in the \unit{mW} power regime.  As derived in Ref.~\cite{wolz2019stimulated},
optimal conditions to transfer population are given when sending equally intense light to stimulate the desired individual $n \rightarrow n-1$ transitions.

\section{\label{s:source}THz sources}

The spectral range in the THz region -- also called far-infrared or sub-mm region, depending on the community (\unit[1]{THz} corresponds to \unit[33]{cm$^{-1}$}, to \unit[$\sim 4$]{meV}, and to a wavelength of \unit[0.3]{mm}) -- is situated at frequencies at which powerful sources are not easily available and \unit[]{mW} power is roughly the bottleneck even if THz technology is a fast moving field (see for instance the reviews given in \cite{Latzel2017,2015JaJAP..54l0101H,dhillon20172017,zhong2017optically,elsasser2019concepts}). The multiple frequencies light necessary for deexcitation or state mixing in antihydrogen can be generated via two techniques:~either via several narrowband sources that emit a sharp spectrum at the wavelengths required to drive single or few transitions, or via a single broadband source that covers the frequency range of all required transitions.
In the following, we will give a general overview of the constraints and limitations of either solution.

A first general constrain, whatever the source is, is linked to the fact that all transitions should be driven simultaneously (and not sequentially) in order to avoid a mere population exchange between the levels. In the following, we will thus restrain ourselves to fixed frequency sources. We first study the possible narrowband sources and then the broadband ones.

\subsection{Narrowband THz sources}
\label{subsec:NarrowBandTHzsource}
In the case of narrowband sources, particular atomic transitions can be targeted and therefore the power provided by the source at those wavelengths is entirely used to drive the transition. Thus, ionization due to off-resonant wavelengths can be minimized. The usage of multiple sources allows to implement the correct power scaling as a function of output frequency increasing the efficiency of the deexcitation. However, when stimulating all $\Delta n = -1$ transitions from $n=30$ down to $n'=5$ a totality of 25 wavelengths is required. In the presence of a magnetic field which leads to significant degeneracy lifting of the levels, the necessary number of (very) narrowband sources can even increase further due to the spectral broadening of the atomic transitions.
In view of the high number of desired wavelengths that need to be produced the usage of expensive direct synthesis such as quantum cascade or molecular lasers is not an option. Furthermore, as mentioned earlier, the exact distribution of quantum states populated during antihydrogen synthesis is not well known and thus a versatile apparatus is needed to adapt the frequencies generated and used for mixing. Given this point, CMOS-based terahertz sources or powerful diodes ($>\unit[1]{mW}$) are not versatile enough solutions, due to the requirement of several frequency multiplications and the necessity of many waveguides given their cut-off frequencies.

In contrast, photomixing or optical rectification \cite{2003OExpr..11.2486A} seems to be an attractive option. Given $n_0$ different laser frequencies $\nu_i$ input signals, the photomixer optical beatnote produces, in the ultra-fast semiconductor material, THz waves at all $\nu_i-\nu_j$ frequencies; the number of which being $n_0(n_0-1)/2$. Photomixing can nowadays reach the $\unit[]{mW}$ level, shared by all generated frequencies \cite{2011JAP...109f1301P}. The $n_0$ laser inputs can be produced using pulse shaping from a single broadband laser source \cite{liu1996terahertz,metcalf2013fully,hamamda2015ro,finneran2015decade}. Given the limitation of a photomixer's total output power, the device is an especially attractive solution for THz mixing purposes (and not necessarily deexcitation towards low $n'$) where the total power is divided up into less frequencies. As mentioned in section \ref{s: requirements}, this is the case for schemes relying on deexcitation lasers. Additionally, the maximum achievable output power rapidly decreases towards the few THz frequency region rendering the device unfit for the deexcitation purpose below $n<15$. We conclude that, in particular for the THz mixing scheme, photomixers exhibit very attractive characteristics.
Furthermore the photomixer simply reproduces the beating in the laser spectrum and can thus also be used as a broadband source.

\subsection{Broadband THz sources}
Using a broadband source has the main advantage that a single device might be able to drive many transitions significantly facilitating the experimental implementation. The obvious drawback, however, is that most of the power will not be emitted at resonant frequencies and thus much higher total power would be required to drive the needed transitions. This might lead to significant losses due to ionization \cite{glukhov2010blackbody,merkt2016} even if filters can be used to reduce this effect. As pointed out earlier, the source output power should ideally be constant over the exploited range of emitted wavelengths which is more difficult to implement with a single broadband source.

Portable synchrotron 
\cite{griffiths2006instrumentation} or table-top Free Electron Laser sources \cite{hooker2013developments,shibata1997broadband,2015PJAB...91..223N,wetzels2006far} would be ideal broadband sources with intense radiance in the far-infrared region, but the costs of such apparatus are still prohibitive. A possible alternative is the use of femtosecond mode-locked lasers to generate very short THz pulses using optical rectification, surface emitters or photoconductive (Auston) switches. However, we can only use sources with fast repetition rates in order for the spontaneous emission to depopulate all levels. Unfortunately, even though  photoconductive switches with $\unit[]{mW}$ THz average output power exist \cite{brown2008milliwatt}, and THz bandwidth in excess of \unit[4]{THz} with power up to $\unit[64]{\mu W}$ as well as optical-power-to-THz-power conversion efficiencies of $\sim 10^{-3}$ have been demonstrated \cite{2013ApPhL.103f1103D}, the efficiency drops to $\sim 10^{-5}$ for fast repetition rates low femtosecond pulse energies. Thus, if using a standard oscillator providing for instance \unit[1]{W} average power, no more than $\unit[10]{\mu W}$ total output power is expected \cite{2015JaJAP..54l0101H}. Such sources have been tested to drive transitions between Rydberg atoms, but with only $10\%$ of the population transfer from the $n=50$ initial states down to $n<40$ 
\cite{mandal2010half,kopyciuk2010deexcitation,takamine2017population}.

A simple solution would consist of a blackbody emitter which  efficiently radiates in the THz range \cite{brundermann2012terahertz}. A \unit[1000]{K} blackbody emits in the far infrared region of \unit[0.1-5]{THz}, with a band radiance of \unit[4]{mW/cm$^2$} which seems perfectly compatible with the requirements found for the antihydrogen deexcitation purpose.
Such a radiation source has been proposed in order to cool internal degrees of freedom of MgH$^+$ molecular ions \cite{2004JPhB...37.4571V}.
Between about $400$ and $\unit[100]{cm^{-1}}$, the radiant power emitted by a silicon carbide (Globar) source is as high as any conventional infrared source, but below $\unit[100]{cm^{-1}}$, as for Nernst lamps or glowers that become transparent below about $\unit[200]{cm^{-1}}$, the emissivity is low. In the region between $\sim \unit[50-10]{cm^{-1}}$ it is thus customary to use a high-pressure mercury lamp  with a spectrum close to a blackbody one of effective temperature of \unit[1000-5000]{K} \cite{griffiths2006instrumentation,brundermann2012terahertz,cann1969light,wolfe1978infrared,1996InPhT..37..471K,buijs2002incandescent,2003PhRvL..90i4801A}.

\section{\label{s:experiment} Experimental caesium test setup}

\begin{figure}
\includegraphics[width=1\linewidth]{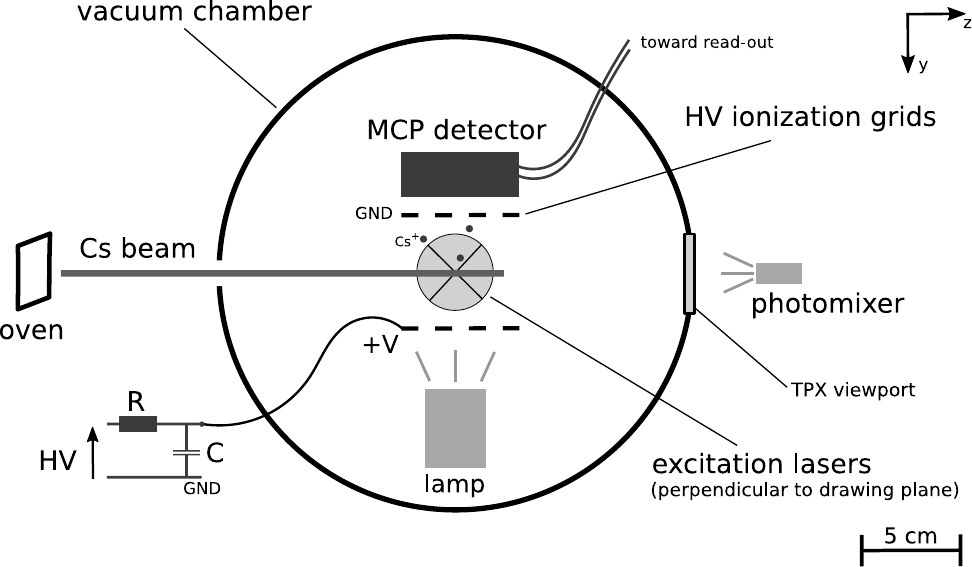}
\caption{Illustration of the experimental caesium beam setup.}
\label{fig:Illustration_setup}
\end{figure}

In order to experimentally assess the potential of the discussed source types, to evaluate realistic power outputs, and to study the suitability of the sources for application to antihydrogen state mixing and deexcitation, we have tested, on a beam of excited Rydberg caesium atoms, the narrow- and broadband solution which seemed most optimal.
The reason to use caesium and not directly hydrogen atoms is mainly due to the fact that, compared to hydrogen, light to manipulate caesium atoms is much easier to generate and off-the-shelf solutions readily exist. However, alkaline Rydberg atoms, such as caesium, have a behavior close to that of hydrogen.

In our experimental setup, illustrated in Fig.~\ref{fig:Illustration_setup}, a caesium effusive beam emitted out of an oven enters a vacuum chamber. The atoms are excited by a cw diode at \unit[852]{nm} from the $6\mathrm{S}_{1/2}$ to the $6\mathrm{P}_{3/2}$ level. A second tunable pulsed laser (OPO pumped by a Nd:YAG) then addresses the $n$S or $n$D Rydberg level. The excitation lasers are sent perpendicular to the beam direction. Two grids opposing each other perpendicular to the beam direction introduce an electric field to field ionize the atoms and study the population of each $(n,l)$ state. 
The THz radiation emitted by a narrowband photomixer outside the chamber can be sent through a THz transparent viewport towards the excited Cs atoms. Alternatively, a broadband lamp is mounted inside the chamber in proximity to the measurement region to stimulate a population transfer

The caesium state population was studied by applying a high voltage pulse to the lower grid (cf.~Fig.~\ref{fig:Illustration_setup}, the other grid was grounded) of the field ionizer surrounding the atomic beam at a given delay time $t_D$ with respect to the laser excitation pulse. The ionizing field was ramped making use of an RC circuit with a rise time of $\unit[4]{\mu s}$. Since each state ionizes at a given electric field strength, the state distribution can be probed by collecting either the ions or electrons from the ionization on a Chevron stack micro-channel plate (MCP) charge detector \cite{ducas1979detection,hollberg1984measurement}.

\begin{figure}
\includegraphics[width=1\linewidth]{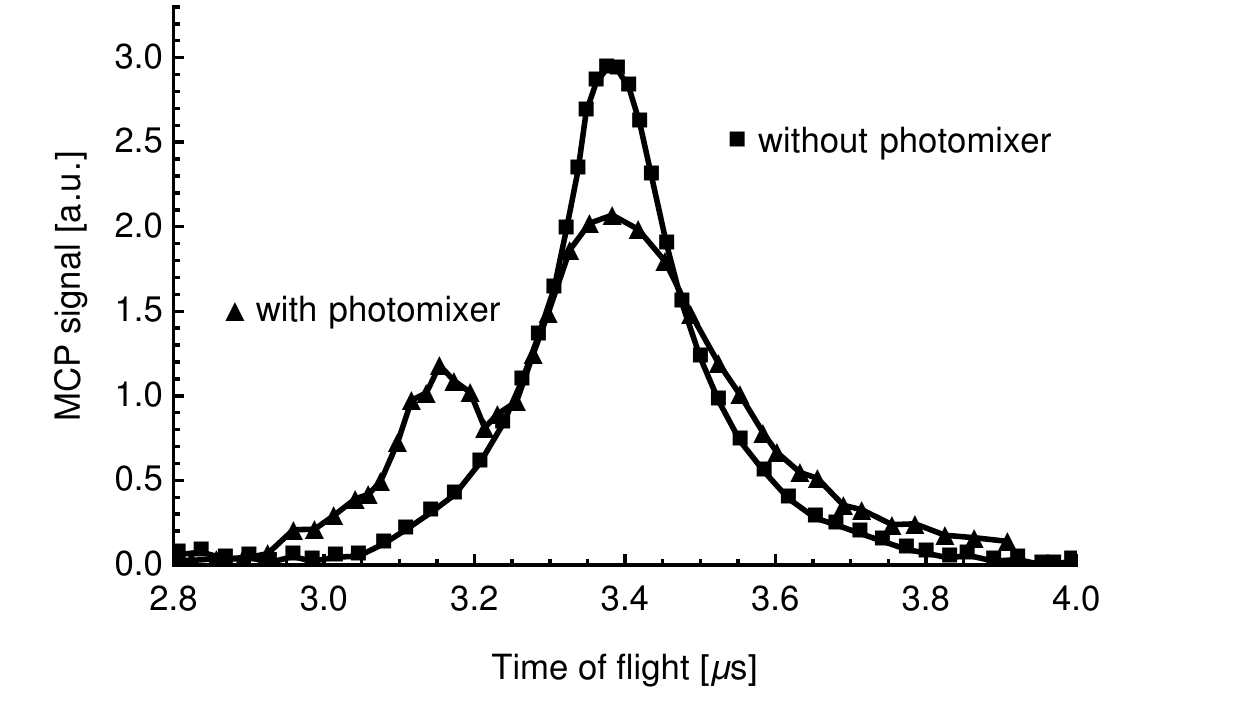}
\caption{Caesium population transfer from the $36\mathrm{S}_{1/2}$ to the $36\mathrm{P}_{3/2}$ level. The obtained MCP signal is plotted for case (1) where the photomixer is switched on (triangle) and case (2) where the photomixer is turned off (square). We indicate on the x-axis the time reference of the signal to the high voltage ramp that is applied to the field ionizer grids. The Cs atoms ionize at a given electric field strength and accelerate towards the MCP. The ionization rate of the $36\mathrm{P}_{3/2}$ level peaks around $\sim \unit[3.15]{\mu s}$ after the high voltage ramp is started. The detection rate of ions originating from the ionization of the $36\mathrm{S}_{1/2}$ level reaches its maximum approximately $\unit[250]{ns}$ later. To improve the readability, the signals are averaged over $\unit[0.4]{\mu s}$}
\label{fig:PhotomixSansAvec}
\end{figure}

We tested a commercial (GaAs Toptica) photomixer acting as a THz source stimulating the $\unit[97]{GHz}$ transition between the initially excited $36 \mathrm{S}_{1/2}$ state towards the $36 \mathrm{P}_{3/2}$ Rydberg state. This transition was chosen due to a strong dipole transition, easy laser excitation and a well defined field ionization signal. Undoubtedly, much more cost-effective, convenient and efficient ways to induce a $\unit[97]{GHz}$ transition would have been to use a voltage-controlled oscillator (VCO), semi-conductor (Gunn or IMPATT diode), backward-wave oscillator or a submillimeter-wave source based on harmonic generation of microwave radiation. However, our goal was not to drive specifically this transition, but to demonstrate the use of a photomixer to drive Rydberg transitions. This technology allows to create a spectrum of multiple sharp frequencies to simultaneously drive many transitions in antihydrogen which ultimately results in a deexcitation of the atoms.
In the context of this proof-of-principle experiment, mixing of near $\unit[852]{nm}$ laser lines from a Ti:Sa laser and a diode laser was used to produce $\sim \unit[1]{\mu W}$ THz output power at $\unit[97]{GHz}$ with a spectral linewidth which reproduces the one of the input lasers ($<\unit[5]{MHz}$). The THz light was sent through a TPX viewport (transparent to THz radiation) and illuminated the sample for $\sim\unit[10]{\mu s}$. A population transfer is clearly visible in Fig.~\ref{fig:PhotomixSansAvec} and amounts to $\sim 15\%$  corresponding to a stimulated transition rate of the order of $\unit[10^4]{s^{-1}}$. In theory, the large Cs dipole matrix element ($554.4 e a_0$ for the $36\mathrm{S}_{1/2} \rightarrow 36\mathrm{P}_{3/2}$ transition \cite{SIBALIC2017319}) should lead to a much faster transition rate of $\Omega \sim \unit[  10^8]{s^{-1}}$ when assuming a light intensity of  $\sim\unit[1]{\mu W/cm^2}$. The experimentally observed lower rate is mainly explained by a transition broadening due to large field inhomogeneities in the region traversed by the Cs beam. Indeed, in our geometry, the MCP produces a fringe field between the two field-ionizer plates that can reach tens of $ \unit[]{V/cm}$ leading to a broadening of tens of GHz for the transition addressed \cite{SIBALIC2017319}. 
This measurement demonstrates the first, to our knowledge, use of a photomixer to stimulate Rydberg transitions in caesium atoms.

\begin{figure}
\includegraphics[width=1\linewidth]{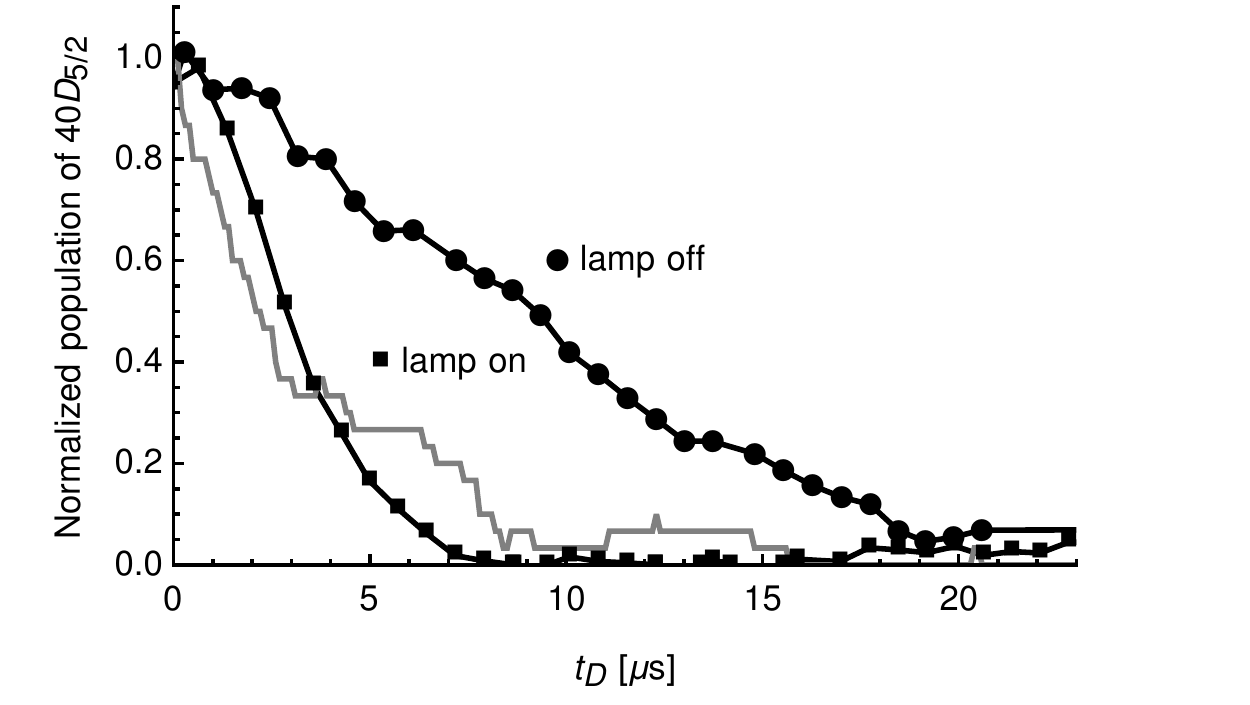}
\caption{Experimentally measured lifetimes of the caesium $40\mathrm{D}_{5/2}$ level with and without a lamp (globar type). The time $t_D$, given on the x-axis, indicates the time delay of the field ionization ramp with respect to the excitation laser. We include simulation results for a \unit[1100]{K} blackbody spectrum (gray). To improve the readability the experimental signals are averaged over $\unit[0.1]{\mu s}$}
\label{fig:ExpResults}
\end{figure}

Fig.~\ref{fig:ExpResults} shows results obtained using a globar type (ASB-IR-12K from Spectral Products) lamp which is a silicon nitride emitter mounted in a $\unit[1]{inch}$ parabolic reflector that is small enough to be placed inside the vacuum chamber $\sim \unit[2]{cm}$ away from where the caesium atoms are excited and ionized. Here, the delay time $t_D$ of the applied ionizing field ramp with respect to the excitation laser was varied to study the population of a given state (that ionizes at a given field strength) as a function of time. To probe the population of these states we integrate the signal in a $\sim \unit[200]{ns}$ time window around the mean arrival time of the field ionization signal. The desired signal  can thus be slightly contaminated by the ionization signal from nearby states. We compare the lifetimes of the state for stimulated population transfer (lamp on) and sole spontaneous emission (lamp off). Fig.~\ref{fig:ExpResults} shows the results obtained for the $40\mathrm{D}_{5/2}$ level. This level was chosen because $n\sim 40$ is close to the highest level that we would hope to transfer in the case of antihydrogen \cite{wolz2019stimulated}. Although the decay curves are non-exponential, we indicate the $1/e$ depopulation time that decreases from $\unit[11]{\mu s}$ to $\unit[3.5]{\mu s}$ using the lamp. To interpret this result, we simulate the spontaneous and light induced depopulation of the $40\mathrm{D}_{5/2}$ state within the caesium atomic system. Dealing with non-coherent light sources we place ourselves in the low saturation limit and reduce the optical Bloch equations to a much simpler set of rate equations. The resulting matrix system is numerically solved for a few hundred atoms as detailed in \cite{2014PhRvA..89d3410C}. The simulations indicate that the enhancement of the decay achieved experimentally is comparable to the simulation result obtained by implementing a light source that emits an isotropic blackbody spectrum of $\sim\unit[1100]{K}$. This is close, and even slightly higher than the temperature emitted by the collimated lamp. Since the device is mounted in close proximity to the caesium beam, it is  possible that the radiative spectrum is indeed focused on the atoms.
In addition, we observed that $\sim 50 \%$ of the atoms are either excited to higher levels or photoionized \cite{vieil2018}. However, we note that filters, such as TPX, PTFE or Teflon \cite{brundermann2012terahertz}, can be used to cut out the low (to avoid $n \rightarrow n+1$ transitions) and high (to avoid direct photoionization) frequency parts of the spectrum that lead to these effects \cite{vieil2018}.

We note that in the cryogenic environment of an antihydrogen experiment the installation of such a high temperature lamp in the vicinity of the atoms remains hypothetical. However, using transport of THz radiation by, for example, a metallic light pipe is simple and efficient \cite{brundermann2012terahertz}. We investigated the transmission efficiency of the lamp's broadband spectrum with a $\unit[30]{cm}$ long copper tube (diameter: $\sim \unit[1]{cm}$) and could transfer $\geq 94.5 \%$ \cite{vieil2018} of the radiation.

\section{Conclusions}
This work reviewed different methods of generating light in the THz region to stimulate the decay of Rydberg antihydrogen atoms towards ground state. 

We commissioned a beamline to study Rydberg population transfer in alkaline atoms. Cesium atoms are much easier to produce than (anti-)hydrogen and are thus ideal for proof-of-principle studies. 
A $\sim 15\%$ population transfer within the $n=36$ manifold was demonstrated using photomixing at $\sim \unit[0.1]{THz}$.
Such THz transitions between Rydberg states can be used to mix the states of antihydrogen atoms while a laser deexcites the Rydberg state distribution towards low $n$-manifolds \cite{wolz2019stimulated}.
Because antihydrogen is formed (by collisions) in many Rydberg levels this mixing and deexcitation requires several frequencies.
The photomixer frequency range of $\sim\unit[0-2]{THz}$ is especially adapted to this purpose. An attractive solution to generate a continuum around the few hundred $\unit[]{GHz}$ region (needed for Rydberg states around $25 \leq n \leq 30$) could be the use of tapered amplifiers, i.~e. semiconductor optical amplifiers, or the amplified spontaneous emission output of an optical amplifier, as radiation inputs towards the photomixer.

A deexcitation of the Cs $40\mathrm{D}_{5/2}$ was observed using a blackbody type light source. However, the ionization fraction for a broadband source lies around $\sim 50 \%$ and is significantly elevated compared to the use of narrowband light sources that emit sharp frequencies targeted towards single $n \rightarrow n-1$ transitions. In combination with the lack of flexibility in the source output power distribution as a function of the emitted wavelength, broadband sources seem consequently rather unfit for the antimatter application. In particular the high ionization potential must be pointed out as a very harmful effect in the context of antihydrogen experiments where atoms are only available in the few hundreds at a time following a complicated synthesis procedure at CERN's Antiproton Decelerator.

We conclude that in particular photomixing has potential for an application in experiments aiming for a deexcitation of antihydrogen atoms. As pointed out, the field of THz light sources is rapidly evolving  and output powers in the \unit{mW} range have been demonstrated. In view of the theoretical studies published in \cite{wolz2019stimulated} and the complementary experimental conclusions drawn in this manuscript, we aim to next demonstrate the first experimental deexcitation of Rydberg hydrogen atoms to ground state in a few tens of microsecond time scales. Addressing this long standing issue in the antihydrogen community has the potential to pave the way for further antimatter precision measurements at CERN.

\section{Acknowledgements}
We dedicate this work to the memory of our co-author Emiliya Dimova who passed away at the age of 50.\newline
This work has been sponsored by the Wolfgang Gentner CERN Doctoral Student Program of the German Federal Ministry of Education and Research (grant no. 05E15CHA, university supervision by Norbert Pietralla).
It was supported by the Studienstiftung des Deutschen Volkes and the Bulgarian Science Fund Grant DN 18/14.

\section{Author contribution statement}
All authors contributed to the work reported in this manuscript.

\bibliographystyle{unsrt}
\bibliography{2020_sep_biblio.bib}

\begin{thebibliography}{10}

\bibitem{AlP_Hbar_Accumulation}
M.~Ahmadi, B.~X.~R. Alves, C.~J. Baker, William Bertsche, et~al.
\newblock Antihydrogen accumulation for fundamental symmetry tests.
\newblock {\em Nature Communications}, 8:681, Dec 2017.

\bibitem{KUR14}
N.~Kuroda, S.~Ulmer, D.~J. Murtagh, S.~Van~Gorp, et~al.
\newblock A source of antihydrogen for in-flight hyperfine spectroscopy.
\newblock {\em Nature Communications}, 5:3089, Jan 2014.

\bibitem{PhysRevLett.108.113002}
G.~Gabrielse, R.~Kalra, W.~S. Kolthammer, R.~McConnell, et~al.
\newblock Trapped antihydrogen in its ground state.
\newblock {\em Phys. Rev. Lett.}, 108:113002, Mar 2012.

\bibitem{Gabrielse2002}
G.~{Gabrielse}, N.~S. {Bowden}, P.~{Oxley}, A.~{Speck}, et~al.
\newblock Background-free observation of cold antihydrogen with
  field-ionization analysis of its states.
\newblock {\em Phys. Rev. Lett.}, 89(21):213401, Oct 2002.

\bibitem{rsa_Mal_18}
C.~Malbrunot, C.~Amsler, S.~Arguedas~Cuendis, H.~Breuker, et~al.
\newblock The {ASACUSA} antihydrogen and hydrogen program: results and
  prospects.
\newblock {\em Philosophical Transactions of the Royal Society A: Mathematical,
  Physical and Engineering Sciences}, 376(2116):20170273, Feb 2018.

\bibitem{ROB08}
F.~Robicheaux.
\newblock Atomic processes in antihydrogen experiments: a theoretical and
  computational perspective.
\newblock {\em Journal of Physics B: Atomic, Molecular and Optical Physics},
  41(19):192001, Sep 2008.

\bibitem{radics2014scaling}
B.~Radics, D.~J. Murtagh, Y.~Yamazaki, and F.~Robicheaux.
\newblock Scaling behavior of the ground-state antihydrogen yield as a function
  of positron density and temperature from classical-trajectory monte carlo
  simulations.
\newblock {\em Phys. Rev. A}, 90:032704, Sep 2014.

\bibitem{Jonsell_2019}
S.~Jonsell and M.~Charlton.
\newblock Formation of antihydrogen beams from positron{\textendash}antiproton
  interactions.
\newblock {\em New Journal of Physics}, 21(7):073020, Jul 2019.

\bibitem{Krasnicky_2019}
D.~{Krasnick{\'y}}, G.~Testera, and N.~Zurlo.
\newblock Comparison of classical and quantum models of anti-hydrogen formation
  through charge exchange.
\newblock {\em Journal of Physics B: Atomic, Molecular and Optical Physics},
  52(11):115202, May 2019.

\bibitem{2016PhRvA..94b2714K}
D.~{Krasnick{\'y}}, R.~{Caravita}, C.~{Canali}, and G.~{Testera}.
\newblock {Cross section for Rydberg antihydrogen production via charge
  exchange between Rydberg positroniums and antiprotons in a magnetic field}.
\newblock {\em Phys. Rev. A}, 94(2):022714, Aug 2016.

\bibitem{2012CQGra..29r4009D}
M.~Doser, C.~Amsler, A.~Belov, G.~Bonomi, et~al.
\newblock {Exploring the WEP with a pulsed cold beam of antihydrogen}.
\newblock {\em Classical and Quantum Gravity}, 29(18):184009, Aug 2012.

\bibitem{PhysRevA.31.495}
Edward~S. Chang.
\newblock Radiative lifetime of hydrogenic and quasihydrogenic atoms.
\newblock {\em Phys. Rev. A}, 31:495--498, Jan 1985.

\bibitem{Topccu2006}
T.~{Top{\c c}u} and F.~{Robicheaux}.
\newblock Radiative cascade of highly excited hydrogen atoms in strong magnetic
  fields.
\newblock {\em Phys. Rev. A}, 73(4):043405, Apr 2006.

\bibitem{wolz2019stimulated}
T.~Wolz, C.~Malbrunot, M.~Vieille-Grosjean, and D.~Comparat.
\newblock Stimulated decay and formation of antihydrogen atoms.
\newblock {\em Phys. Rev. A}, 101:043412, Apr 2020.

\bibitem{gallagher_1994}
T.~F. Gallagher.
\newblock {\em Rydberg Atoms}.
\newblock Cambridge Monographs on Atomic, Molecular and Chemical Physics.
  Cambridge University Press, 1994.

\bibitem{COM181}
D.~Comparat and C.~Malbrunot.
\newblock Laser stimulated deexcitation of {R}ydberg antihydrogen atoms.
\newblock {\em Phys. Rev. A}, 99:013418, Jan 2019.

\bibitem{Latzel2017}
P.~Latzel, F.~Pavanello, S.~Bretin, M.~Billet, et~al.
\newblock High efficiency {UTC} photodiode for high spectral efficiency {THz}
  links.
\newblock In {\em 2017 42nd International Conference on Infrared, Millimeter,
  and Terahertz Waves (IRMMW-THz)}, pages 1--2, Aug 2017.

\bibitem{2015JaJAP..54l0101H}
M.~Hangyo.
\newblock {Development and future prospects of terahertz technology}.
\newblock {\em Japanese Journal of Applied Physics}, 54(12):120101, Dec 2015.

\bibitem{dhillon20172017}
S.~S. Dhillon, M.~S. Vitiello, E.~H. Linfield, A.G. Davies, et~al.
\newblock The 2017 terahertz science and technology roadmap.
\newblock {\em Journal of Physics D: Applied Physics}, 50(4):043001, Jan 2017.

\bibitem{zhong2017optically}
K.~Zhong, W.~Shi, D.~Xu, P.~Liu, et~al.
\newblock Optically pumped terahertz sources.
\newblock {\em Science China Technological Sciences}, 60(12):1801--1818, Jun
  2017.

\bibitem{elsasser2019concepts}
T.~Els{\"a}sser, K.~Reimann, and M.~Woerner.
\newblock {\em Concepts and Applications of Nonlinear Terahertz Spectroscopy}.
\newblock Morgan \& Claypool Publishers, Feb 2019.

\bibitem{2003OExpr..11.2486A}
J.~Ahn, A.~V. Efimov, R.~D. Averitt, and A.~T. Taylor.
\newblock {Terahertz waveform synthesis via optical rectification of shaped
  ultrafast laser pulses}.
\newblock {\em Optics Express}, 11:2486, Oct 2003.

\bibitem{2011JAP...109f1301P}
S.~Preu, G.~H. D{\"o}hler, S.~Malzer, L.~J. Wang, et~al.
\newblock {Tunable, continuous-wave Terahertz photomixer sources and
  applications}.
\newblock {\em Journal of Applied Physics}, 109(6):061301--061301, Mar 2011.

\bibitem{liu1996terahertz}
Y.~Liu, S.~Park, and A.~M. Weiner.
\newblock Terahertz waveform synthesis via optical pulse shaping.
\newblock {\em Selected Topics in Quantum Electronics, IEEE Journal of},
  2(3):709--719, Sep 1996.

\bibitem{metcalf2013fully}
A.~J. Metcalf, V.~R. Supradeepa, D.~E. Leaird, A.~M. Weiner, et~al.
\newblock Fully programmable two-dimensional pulse shaper for broadband
  line-by-line amplitude and phase control.
\newblock {\em Optics express}, 21(23):28029--28039, Nov 2013.

\bibitem{hamamda2015ro}
M.~Hamamda, P.~Pillet, H.~Lignier, and D.~Comparat.
\newblock Ro-vibrational cooling of molecules and prospects.
\newblock {\em Journal of Physics B: Atomic, Molecular and Optical Physics},
  48(18):182001, Aug 2015.

\bibitem{finneran2015decade}
A.~I. Finneran, J.~T. Good, D.~B. Holland, P.~B. Carroll, et~al.
\newblock Decade-spanning high-precision terahertz frequency comb.
\newblock {\em Phys. Rev. Lett.}, 114(16):163902, Apr 2015.

\bibitem{glukhov2010blackbody}
I.~L. Glukhov, E.~A. Nekipelov, and V.~D. Ovsiannikov.
\newblock Blackbody-induced decay, excitation and ionization rates for rydberg
  states in hydrogen and helium atoms.
\newblock {\em Journal of Physics B: Atomic, Molecular and Optical Physics},
  43(12):125002, Jun 2010.

\bibitem{merkt2016}
C.~Seiler, J.~A. Agner, P.~Pillet, and F.~Merkt.
\newblock Radiative and collisional processes in translationally cold samples
  of hydrogen rydberg atoms studied in an electrostatic trap.
\newblock {\em Journal of Physics B: Atomic, Molecular and Optical Physics},
  49(9):094006, Apr 2016.

\bibitem{griffiths2006instrumentation}
P.~R. Griffiths and C.~C. Homes.
\newblock {\em Instrumentation for Far-Infrared Spectroscopy}.
\newblock Wiley Online Library, 2006.

\bibitem{hooker2013developments}
S.~M. Hooker.
\newblock Developments in laser-driven plasma accelerators.
\newblock {\em Nature Photonics}, 7(10):775--782, Sep 2013.

\bibitem{shibata1997broadband}
Y.~Shibata, K.~Ishi, S.~Ono, Y.~Inoue, et~al.
\newblock Broadband free electron laser by the use of prebunched electron beam.
\newblock {\em Phys. Rev. Lett.}, 78(14):2740, Apr 1997.

\bibitem{2015PJAB...91..223N}
K.~{Nakajima}.
\newblock {Laser-driven electron beam and radiation sources for basic, medical
  and industrial sciences}.
\newblock {\em Proceeding of the Japan Academy, Series B}, 91:223--245, 2015.

\bibitem{wetzels2006far}
A.~Wetzels, A.~G{\"u}rtler, L.~D. Noordam, and F.~Robicheaux.
\newblock Far-infrared {Rydberg-Rydberg} transitions in a magnetic field:
  Deexcitation of antihydrogen atoms.
\newblock {\em Phys. Rev. A}, 73(6):062507, Jun 2006.

\bibitem{brown2008milliwatt}
E.~R. Brown.
\newblock Milliwatt thz average output power from a photoconductive switch.
\newblock In {\em {Infrared, Millimeter and Terahertz Waves, 2008. IRMMW-THz
  2008. 33rd International Conference on}}, pages 1--2. IEEE, 2008.

\bibitem{2013ApPhL.103f1103D}
R.~J.~B. {Dietz}, B.~{Globisch}, M.~{Gerhard}, A.~{Velauthapillai}, et~al.
\newblock {64 {$\mu$}W pulsed terahertz emission from growth optimized
  InGaAs/InAlAs heterostructures with separated photoconductive and trapping
  regions}.
\newblock {\em Applied Physics Letters}, 103(6):061103, Aug 2013.

\bibitem{mandal2010half}
P.~K. Mandal and A.~Speck.
\newblock Half-cycle-pulse-train induced state redistribution of rydberg atoms.
\newblock {\em Phys. Rev. A}, 81(1):013401, Oct 2010.

\bibitem{kopyciuk2010deexcitation}
T.~Kopyciuk.
\newblock Deexcitation of one-dimensional {Rydberg} atoms with a chirped train
  of half-cycle pulses.
\newblock {\em Physics Letters A}, 374(34):3464--3467, Jul 2010.

\bibitem{takamine2017population}
A.~Takamine, R.~Shiozuka, and H.~Maeda.
\newblock Population redistribution of cold rydberg atoms.
\newblock In {\em Proceedings of the 12th International Conference on Low
  Energy Antiproton Physics (LEAP2016)}, page 011025, Nov 2017.

\bibitem{brundermann2012terahertz}
E.~Br{\"u}ndermann, H.~H{\"u}bers, and M.~F. Kimmitt.
\newblock {\em Terahertz Techniques}, volume 151.
\newblock Springer, 2012.

\bibitem{2004JPhB...37.4571V}
I.~S. Vogelius, L.~B. Madsen, and M.~Drewsen.
\newblock {Rotational cooling of molecules using lamps}.
\newblock {\em Journal of Physics B}, 37:4571--4574, Nov 2004.

\bibitem{cann1969light}
M.~W.~P. Cann.
\newblock Light sources in the 0.15--20-$\mu$ spectral range.
\newblock {\em Applied optics}, 8(8):1645--1661, Aug 1969.

\bibitem{wolfe1978infrared}
W.~L. Wolfe and G.~J. Zissis.
\newblock {\em The infrared handbook}, volume~1.
\newblock Spie Press, 1978.

\bibitem{1996InPhT..37..471K}
M.~F. {Kimmitt}, J.~E. {Walsh}, C.~L. {Platt}, K.~{Miller}, et~al.
\newblock {Infrared output from a compact high pressure arc source}.
\newblock {\em Infrared Physics and Technology}, 37:471--477, Jun 1996.

\bibitem{buijs2002incandescent}
H.~Buijs.
\newblock {\em Incandescent Sources for Mid-and Far-Infrared Spectrometry}.
\newblock Wiley Online Library, 2002.

\bibitem{2003PhRvL..90i4801A}
M.~Abo-Bakr, J.~Feikes, K.~Holldack, P.~Kuske, et~al.
\newblock {Brilliant, Coherent Far-Infrared (THz) Synchrotron Radiation}.
\newblock {\em Phys. Rev. Lett.}, 90(9):094801--+, Mar 2003.

\bibitem{ducas1979detection}
T.~W. Ducas, W.~P. Spencer, A~G. Vaidyanathan, W.~H. Hamilton, et~al.
\newblock Detection of far-infrared radiation using rydberg atoms.
\newblock {\em Appl. Phys. Lett.}, 35(5):382--384, Aug 1979.

\bibitem{hollberg1984measurement}
L~Hollberg and JL~Hall.
\newblock Measurement of the shift of {Rydberg} energy levels induced by
  blackbody radiation.
\newblock {\em Phys. Rev. Lett.}, 53(3):230, Jul 1984.

\bibitem{SIBALIC2017319}
N.~\^Sibali\'c, J.~D. Pritchard, Adams~C. S., and K.~J. Weatherill.
\newblock {ARC: An open-source library for calculating properties of alkali
  Rydberg atoms}.
\newblock {\em Computer Physics Communications}, 220:319 -- 331, 2017.

\bibitem{2014PhRvA..89d3410C}
D.~Comparat.
\newblock {Molecular cooling via Sisyphus processes}.
\newblock {\em Phys. Rev. A}, 89(4):43410, Apr 2014.

\bibitem{vieil2018}
M.~Vieille~Grosjean.
\newblock {\em Atomes de Rydberg : Etude pour la production d'une source
  d'\'electrons monocin\'etique. D\'esexcitation par radiation THz pour
  l'antihydrog\`ene}.
\newblock PhD thesis, Laboratoire Aim\'{e} Cotton, Orsay, France, 2018.

\end{thebibliography}

\end{document}